\documentclass{ws-procs9x6-cpt25}
\begin{document}

\newcommand{\refeq}[1]{(\ref{#1})}
\def\etal {{\it et al.}}

\title{Full Single-Quantum Control of Particles in Penning Traps for Symmetry Tests at the Quantum Limit}

\author{J.M.\ Cornejo,$^{1,2,a}$,
J.-A.\ Coenders,$^{1,2}$
A.\ Lissel,$^{2}$
N.\ Poljakov,$^{1,2}$
M.\ Prasse,$^{1,2}$
Y.\ Priewich,$^{1,2}$
J.\ Schaper,$^{1,2}$
M.\ Schubert,$^{3}$
B.\ Hampel,$^{3}$
M.\ Schilling,$^{3}$
S.\ Ulmer,$^{4,5}$
and C.\ Ospelkaus$^{1,2,6}$\\}

\address{$^1$Institut  f{\"u}r Quantenoptik, Leibniz Universit{\"a}t Hannover,\\ Welfengarten 1, 30167 Hannover, Germany}

\address{$^2$Laboratorium für Nano- und Quantenengineering, Leibniz Universit{\"a}t Hannover,\\ Schneiderberg 39, 30167 Hannover, Germany}

\address{$^3$Institut für Elektrische Messtechnik und Grundlagen der Elektrotechnik, TU Braunschweig,\\ Hans-Sommer-Straße 66, 38106 Braunschweig, Germany}

\address{$^4$Heinrich-Heine-Universität Düsseldorf,\\
 Universitätsstraße 1, 40225 Düsseldorf, Germany}

\address{$^5$RIKEN, Ulmer Fundamental Symmetries Laboratory,\\
2-1 Hirosawa, Wako, Saitama, 351-0198, Japan}

\address{$^6$Physikalisch-Technische Bundesanstalt,\\ Bundesallee 100, 38116 Braunschweig, Germany}

\address{$^{a}$ Present address: Departamento de Física de la Materia Condensada,\\ Universidad de Cádiz, C. República Saharaui, 11519 Puerto Real, Cádiz, España}

\begin{abstract}

The BASE collaboration aims to measure antimatter systems with the highest precision in order to perform a rigorous test of CPT symmetry and search for physics beyond the Standard Model. As part of the BASE collaboration, we pursue the development of quantum logic inspired cooling and detection techniques for \textit{g}-factor measurements of \mbox{(anti-)protons}. Implementing these methods requires full quantum-level control of individual antimatter particles confined in cryogenic Penning traps. By mapping the (anti-)proton’s internal state onto a co-trapped $^9$Be$^+$ “logic” ion via free Coulomb coupling in a double-well potential, we can accelerate measurement cycles and push \textit{g}-factor precision measurements on \mbox{(anti-)protons} toward the quantum limit. Here, we present an overview of the proposed method and the current status of the project, with special emphasis on the new cryogenic multi-Penning-trap stack and the proton detection system.
 
\end{abstract}

\bodymatter

\section{Introduction}\label{aba:sec1}

The CPT theorem states that a relativistic Lorentz-symmetric quantum field theory must also be CPT invariant.\cite{Pauli} Although individual violations of C, P, or T symmetries are observed in certain weak interactions, the theorem guarantees that the combination of charge conjugation (C), parity inversion (P), and time reversal (T) remains an exact symmetry of all local, Lorentz-invariant quantum field theories.\cite{Ralf} This profound result not only underpins the consistency of particle–antiparticle equivalence but also provides critical constraints on possible extensions to the Standard Model.\cite{datatables} Ultra-high precision comparisons of protons and antiprotons for the stringent test of CPT symmetry in the baryonic sector are pursued in the BASE (Baryon Antibaryon Symmetry Experiment) collaboration.\cite{Smorra2015} To achieve this, advanced cryogenic Penning-trap systems are used to measure the \textit{g}-factor and the charge-to-mass ratio of protons and antiprotons to the highest precision. In this device, single particles are isolated in an ultra-high vacuum environment, where a strong magnetic field provides radial confinement, while static voltages applied to the trap electrodes ensure axial confinement. To date, the BASE collaboration has measured the \textit{g}-factor with an unprecedented precision of 1.5 parts-per-billion (ppb) for the antiproton\cite{Smorra2017} and 0.3~ppb for the proton.\cite{Schneider2017} Additionally, it has carried out the most precise CPT test in the baryonic sector by comparing the proton and antiproton charge-to-mass ratios with a precision of 16 parts-per-trillion.\cite{Borchert2022} While the efficacy of these measurements is evident, current techniques continue to face challenges due to the finite temperature of the particles and the long preparation times that the measurement protocol entails. In this work, we report on the status to experimentally implement quantum logic inspired cooling and detection techniques in order to boost current \textit{g}-factor measurements of \mbox{(anti-)protons} within the BASE collaboration.

\section{Quantum logic inspired techniques for $\bm{g}$-factor measurements with \mbox{(anti-)protons}}\label{aba:sec1}

In recent decades, laser-based techniques have undergone substantial development for the manipulation of atomic ions.\cite{Wineland1998} The application of this concept to other quantum systems\cite{Heinzen1990} has ultimately resulted in the experimental realization of quantum logic spectroscopy,\cite{Schmidt2005} with application to molecular ions,\cite{Wolf2016} atomic clocks,\cite{Rosenband2010} and highly charged ions.\cite{Micke2020} Its use to measure the \textit{g}-factor of \mbox{(anti-)protons} was introduced in Ref.\ [\refcite{Wineland1998}] and it has been comprehensively discussed in Ref.\ [\refcite{Cornejo2021}]. The main difference in its application to antimatter lies in the use of Penning traps instead of radio-frequency traps, which have been used for quantum logic spectroscopy so far.\cite{Schmidt2005,Wolf2016,Rosenband2010,Micke2020} Although recent years have seen growing interest in the application of laser-based techniques in Penning traps,\cite{Goodwin2016,Jain2024,Jordan2019,Bohman2021,Dominguez2018,Cornejo2016} it has not been widely implemented in such devices due to the large Zeeman splittings of the electronic levels caused by the high magnetic fields,\cite{Guti2019,Mielke2021,Paschke2019} as well as the complex cooling dynamics arising from the inherent instability of the magnetron motion.\cite{Thompson2000,Cornejo2023} This motion is one of the three characteristic motions of a particle in a Penning trap, along with the axial and reduced cyclotron motions. These three motions are related to the cyclotron motion through the invariance theorem.\cite{Brown1982} The \textit{g}-factor of an \mbox{(anti-)proton} in a Penning trap can be obtained by determining the cyclotron frequency $\omega_c$ and the spin precession Larmor frequency $\omega_L$ according to:
\begin{equation}
\frac{g}{2}=\frac{\omega_L}{\omega_C}.
\label{aba:eq1}
\end{equation}

The determination of the Larmor frequency of single \mbox{(anti-)protons} in a Penning trap using quantum logic spectroscopy involves a single $^9$Be$^+$ ion as a coolant and detection qubit. First, the $^9$Be$^+$ ion is cooled to its motional ground state via laser cooling and coupled to a single proton or antiproton through the free Coulomb interaction in a double-well potential.\cite{Wineland1998,Heinzen1990} When both ions oscillate at the same frequency, energy is exchanged between them.\cite{Brown2011} Using this technique, a single proton or antiproton can be cooled to its motional ground state. Once both ions are initialized in their motional ground states, the spin state of the \mbox{(anti-)proton} is probed by applying different excitation frequencies around the expected Larmor frequency. For each interrogation, the spin state of the \mbox{(anti-)proton} is first transferred onto its motional mode via spin-motion coupling.\cite{Nitzschke2020} Depending on the spin orientation, a quantum of motion is added to the \mbox{(anti-)proton}, which is subsequently transferred to the $^9$Be$^+$ ion through the Coulomb interaction. Finally, the motional state of the $^9$Be$^+$ ion is detected using laser sideband transitions, allowing the determination of the initial spin state of the \mbox{(anti-)proton}. 

In order to implement this experimental approach, we have constructed and brought into operation a Penning trap system that includes a 5\,T superconducting magnet and a cryogenic mechanical structure designed to provide laser access to the trap stack for manipulation of $^9$Be$^+$ ions.\cite{Niemann2019} Additionally, we have already demonstrated control of the cooling ion,\cite{Niemann2019,Cornejo2023} a novel laser system for Raman transitions,\cite{Mielke2021} sideband spectroscopy,\cite{Cornejo2023} fast adiabatic transport,\cite{Meiners2024,Boehn2025} and also motional ground-state cooling of a single $^9$Be$^+$ ion.\cite{Cornejo2024} The next step is to demonstrate energy exchange between pairs of $^9$Be$^+$ ions,\cite{Meiners2018} and subsequently with a $^9$Be$^+$ ion and a single proton.\cite{Cornejo2021} For this, we have modified our cryogenic Penning trap system, as described below.

\section{Penning trap systems}\label{aba:sec1}

\begin{figure}
\centering
\includegraphics[width=4.3in]{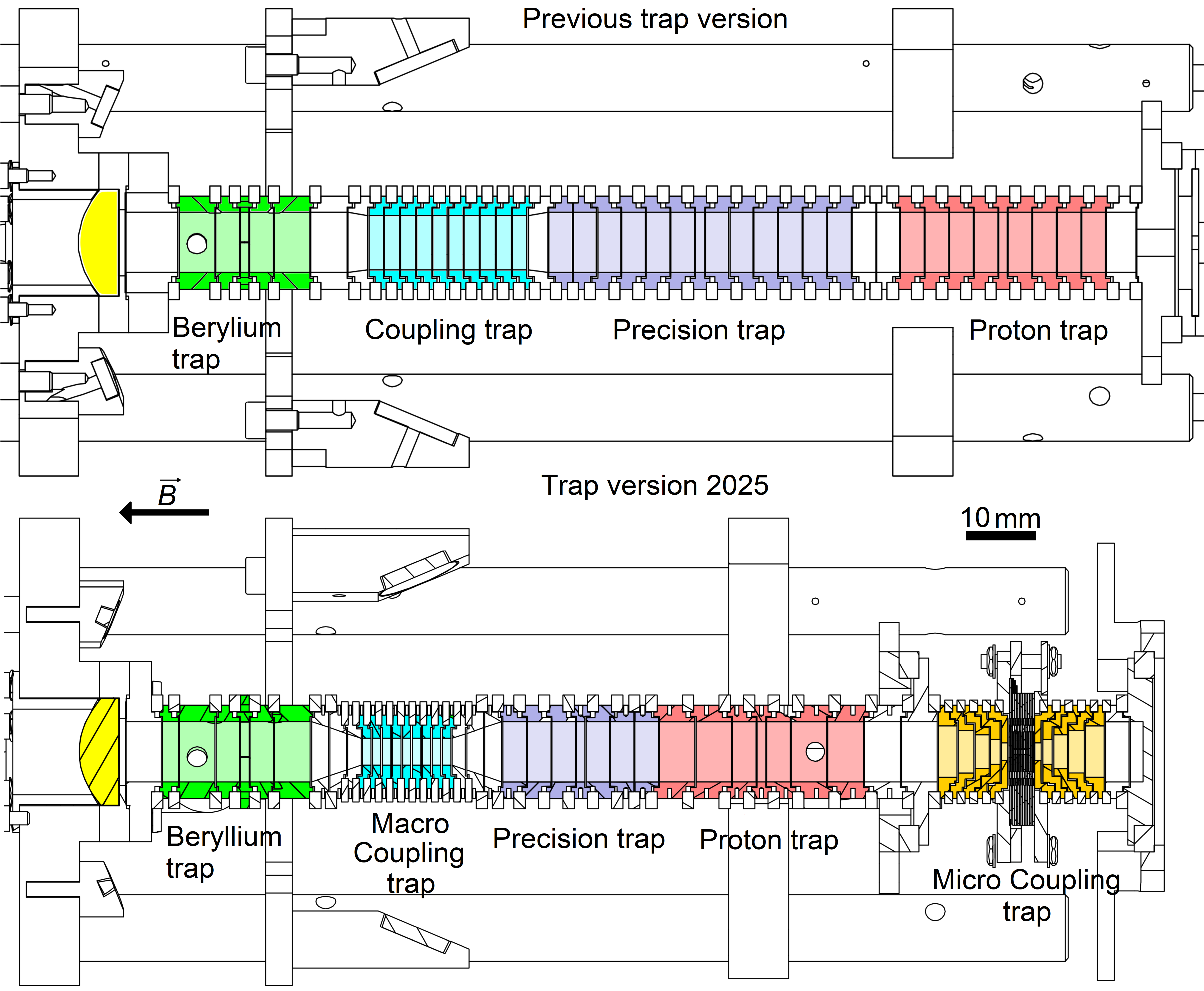}
\caption{Two-dimensional CAD drawing of the complete Penning trap stacks with specialized trap zones. The previous version of the trap is shown in the top panel. Several modifications have been implemented in the updated trap system to enable the manipulation of single protons. The electrodes are made of gold-plated oxygen-free high thermal conductivity (OFHC) copper and are electrically insulated by sapphire rings. The trap is located at the center of a 5\,T superconducting magnet and cooled to a temperature of approximately 5\,K. Further details are provided in the text.}
\label{aba:fig1}
\end{figure}

Figure \ref{aba:fig1} presents a schematic of our Penning trap systems. This setup includes a beryllium trap for laser manipulation of single $^9$Be$^+$ ions, a coupling trap designed to study the coupling between two $^9$Be$^+$ ions, as a preparatory step for investigating the coupling between a $^9$Be$^+$ ion and a single proton, a precision trap for the manipulation of individual protons, a proton trap for the production of single protons, and a micro coupling trap specifically intended for coupling a $^9$Be$^+$ ion with a single proton. Several features have been improved compared to our previous setup (see Fig.\ \ref{aba:fig1}), in which single-ion manipulation of $^9$Be$^+$ ions was already demonstrated. These improvements include modifications to the beryllium trap, the development of new coupling traps, including advancements in microfabrication techniques for the construction of the micro coupling trap, and upgrades to the precision and proton traps for proton production, initialization, and detection.

Ion-ion coupling via free Coulomb interaction is achieved in a specific region of our trap stack by generating a double-well potential. In this scenario, the energy exchange rate is given by
\begin{equation}
\Omega_{ex} = \frac{q_a q_b}{\pi \epsilon_0 s_0^3 \sqrt{m_a m_b} \sqrt{\omega_a \omega_b}},
\label{eqn:rate}
\end{equation}
where $s_0$ is the distance between the ions, $\epsilon_0$ is the vacuum permittivity, and $m_i$, $q_i$, and $\omega_i$ are the mass, charge, and oscillation frequency, respectively, of particle $i \in \{a, b\}$. In our previous Penning trap system,\cite{Cornejo2023, Boehn2025} several attempts were made to study energy exchange between pairs of $^9$Be$^+$ ions, albeit without success. We attribute this mainly to the lack of proper temperature initialization of the ions prior to each coupling attempt. This limitation arose from the inability to transport two different ions to the beryllium trap for laser manipulation because only the most recently produced ion could be transferred. In order to overcome this issue, an additional electrode was incorporated into the beryllium trap, enabling laser manipulation of two ions and reliable temperature initialization prior to each ion coupling attempt. Moreover, the four-segment ring electrode of the beryllium trap has been connected to independent DC voltage supplies in order to compensate for small deviations between the radial AC and DC potentials. With this improvement, we expect to Doppler-cool single $^9$Be$^+$ ions to the Doppler limit temperature of 500\,\textmu K, which is a factor of three lower than that achieved with our previous system.\cite{Cornejo2023} This enhancement will facilitate ground-state cooling and reduce the overall cooling times.    

\begin{figure}
\centering
\includegraphics[width=4.3in]{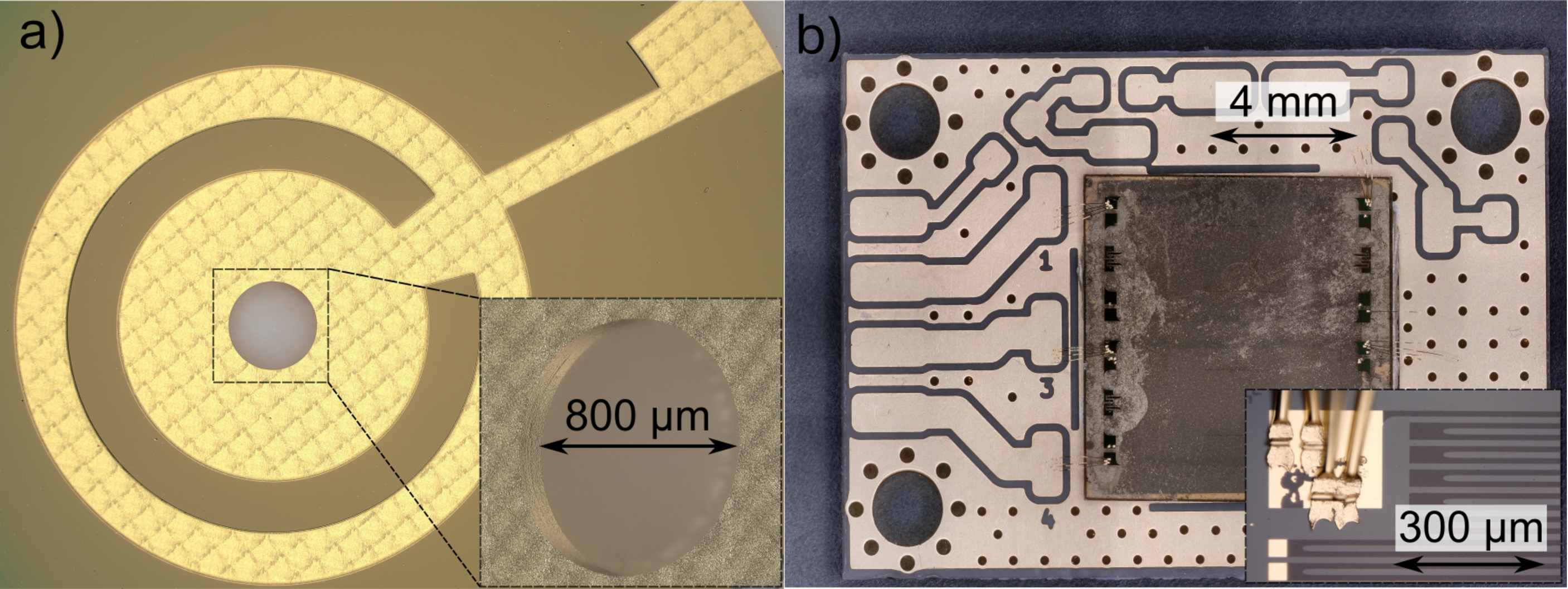}
\caption{a) Photograph of a single electrode from the micro coupling trap, showing the C-shaped structure designed for self-alignment. A detailed view of the inner hole used as a hollow cylindrical electrode, with an inner diameter of 800\,\textmu m, is shown in the bottom right corner. b) Photograph of the Printed Circuit Board (PCB) featuring four kinetic inductance resonators to be tested in our trap system. In the bottom right corner, a microscopic view of a meander structure with gold bonding is shown.}
\label{aba:fig2}
\end{figure}

In addition, we have introduced a smaller macro coupling trap. According to Eq.~(\ref{eqn:rate}), a shorter distance between the ions—i.e., between the two potential minima generated in the double-well potential—leads to higher exchange rates. We have reduced the inner diameter of the trap by a factor of two, resulting in an eightfold increase in the coupling rate under the same conditions. This higher coupling rate facilitates energy exchange between two $^9$Be$^+$ ions and also enables the study of coupling between a single $^9$Be$^+$ ion and a proton. However, in the latter case, the diameter of the macro coupling trap is still too large to produce a double-well potential with a sufficiently small separation between the two potential minima, and thus a sufficiently large exchange rate. To overcome this limitation, a micro coupling trap has been developed, with an inner diameter reduced by a factor of five compared to the macro coupling trap. Under the same coupling conditions, this results in a 125-fold increase in the exchange rate. However, gold-plated copper electrodes cannot be used due to the small dimensions of this micro trap. Instead, it must be fabricated using microfabrication techniques, in a manner similar to that of planar surface radio frequency ion traps\cite{Seidelin2006} but adapted to a cylindrical geometry suitable for high-precision Penning trap experiments. This three-dimensional micro trap consists of 15 disc-shaped electrodes made of structured fused silica, partially gold-plated, with an outer diameter of 22\,mm. It was designed as a self-aligning structure, where each electrode fits precisely into the next in only one specific orientation. The hollow cylindrical electrodes have an inner diameter of 800\,\textmu m and a thickness of 200\,\textmu m. Figure~\ref{aba:fig2}a shows a photo of a single electrode. Additionally, for proper ion transport through the micro trap, the magnetic field lines of the superconducting magnet must be aligned with the trap axis. To achieve this, we are currently developing an alignment system capable of slightly adjusting the position of the entire cryomechanical structure, allowing precise alignment of the trap axis with the magnetic field. Furthermore, several electrodes with progressively reduced inner diameters are used to guide the ions adiabatically into the coupling region, as shown in Fig.~\ref{aba:fig1}. Due to the novel and technically challenging task of transporting ions through the micro trap, in this first implementation, the micro trap has been positioned at one end of the trap stack. This configuration allows experiments with both protons and beryllium ions without obstructing the rest of the trap system. However, in future iterations, the micro trap will be placed between the beryllium trap and the precision trap to enable optimized coupling experiments.

For proton production, we plan to use an ablation laser and a tantalum target located at the proton trap, replacing the electron gun used in our previous setup. Tantalum is an effective adsorber of hydrogen, making it a suitable material for proton production via laser ablation.\cite{Velardi2014} Additionally, the precision trap has been shortened to accommodate the micro coupling trap and to avoid an excessive number of electrodes. For proton detection and thermalization, we will use axial and cyclotron resonators, as in our previous setup. These systems are based on resonant LC circuits, in which an inductor $L$ is connected to a trap electrode. Due to the parasitic capacitance of the system, the circuit behaves as a resonant oscillator with a resonance frequency given by $\omega_{LC} = 1/\sqrt{LC}$. For the axial and cyclotron resonators, the resonance frequencies are matched to the axial and modified cyclotron oscillation frequencies of a single proton in the precision trap. This configuration allows both the detection of image currents induced in the trap electrodes by the proton, and the thermalization of the proton to the temperature of the resonator. However, due to the significant difference in oscillation frequencies—approximately 600\,kHz for the axial motion and 76\,MHz for the reduced cyclotron motion—the two resonators require different designs. The axial resonator is made from superconducting NbTi cable, which provides a high quality factor $Q$. This enables the resonator to be placed at a considerable distance from the trap electrodes while maintaining sufficient sensitivity. In contrast, the cyclotron resonator is made from OFHC copper, which exhibits a lower $Q$ value, requiring it to be located close to the trap electrodes for effective signal detection. In the updated trap design, this presents a challenge due to the physical size of the cyclotron resonator—several centimeters—as well as the need to accommodate mirrors and laser beam paths within the limited space. In order to overcome this limitation, we are also working on the implementation of a kinetic inductance-based resonator. Kinetic inductance arises from the inertial mass of mobile charge carriers in alternating electric fields, manifesting as an equivalent series inductance. This effect is particularly pronounced in conductors with high charge-carrier mobility, such as superconductors.\cite{Meservey1969} As inductance, we use thin films of high-temperature superconducting yttrium barium copper oxide (YBa$_2$Cu$_3$O$_{7-x}$, YBaCuO) over a (LaAlO3)0.3(Sr2TaAlO6)0.7 (LSAT) substrate. Figure~\ref{aba:fig2}b shows the PCB with the substrate and the meander structure formed by the films. The YBaCuO thin films are deposited onto the LSAT substrate via pulsed laser deposition using polycrystalline targets.\cite{Malik2012}

\section{Summary and outlook}

In this work, we have discussed quantum logic inspired techniques for measuring the \textit{g}-factor of \mbox{(anti-)protons} as a stringent test of CPT symmetry in the baryonic sector. Furthermore, we have presented an updated Penning trap system incorporating new methods for proton production and detection, as well as for implementing energy exchange between two $^9$Be$^+$ ions, and subsequently between a single $^9$Be$^+$ ion and a single proton. This is a key requirement for achieving quantum-level control of single protons. Alongside sideband spectroscopy of single protons, these represent the essential techniques to be developed. Additionally, due to advancements in the antiproton delivery system from CERN,\cite{Leonhardt2025} these methods could also be applied to antiprotons within the BASE collaboration.

\section*{Acknowledgments}
This work was supported by PTB, LUH, and DFG through the clusters of excellence QUEST and QuantumFrontiers as well as through the Collaborative Research Center SFB1227 (DQ-mat Project-ID 274200144) and ERC StG ``QLEDS.'' We also acknowledge financial support from the RIKEN Pioneering Project Funding and the MPG/RIKEN/PTB Center for Time, Constants and Fundamental Symmetries.

\end{document}